\documentclass[msmath,amssymb,aps,pra,twocolumn,epsfig,showpacs,bibliography,
lengthcheck,superscriptaddress]{revtex4-1}

\usepackage{graphicx}
\usepackage{dcolumn}
\usepackage{bm}
\usepackage{hyperref}
\usepackage{epstopdf}
\usepackage{subcaption}
\usepackage{caption}
\usepackage{footmisc}
\usepackage{amsmath}
\usepackage{color}
\begin{document}
\title{Herriott-Cavity-Assisted Closed-Loop Xe Isotope Comagnetometer}%
\author{C.-P. Hao}
\author{Q.-Q. Yu}
\author{C.-Q. Yuan}
\author{S.-Q. Liu}
\author{D. Sheng}
\email{dsheng@ustc.edu.cn}
\affiliation{Hefei National Laboratory of Physical Sciences at the Microscale, University of Science and Technology of China, Hefei 230026, China}
\affiliation{Department of Precision Machinery and Precision Instrumentation, Key Laboratory of Precision Scientific Instrumentation of Anhui Higher Education Institutes, University of Science and Technology of China, Hefei 230027, China}

\begin{abstract}
We present in this paper a Herriott-cavity-assisted closed-loop Xe isotope gas comagnetometer. In this system, $^{129}$Xe and $^{131}$Xe atoms are pumped and probed by polarized Rb atoms, and continuously driven by oscillating magnetic fields, whose frequencies are kept on resonance by phase-locked loops (PLLs). Different from other schemes, we use a Herriott cavity to improve the Rb magnetometer sensitivity instead of the parametric modulation method, and this passive method is aimed to improve the system stability while maintaining the sensitivity. This system has demonstrated an angle random walk (ARW) of 0.06~$^\circ$/h$^{1/2}$, and a bias instability of 0.2~$^\circ$/h (0.15~$\mu$Hz) with a bandwidth of 1.5~Hz. By adding a closed-loop Rb isotope comagnetometer, we can extend this system to dual simultaneously working  comagnetometers sharing the same cell. This extended system has wide applications in precision measurements, where we can simultaneously and independently measure the coupling of anomalous fields with proton spin and neutron spin.
\end{abstract}

\maketitle

\section{Introduction}
Atomic comagnetometers use two species of atoms in the same place to simultaneously measure and cancel the magnetic field fluctuations~\cite{lamoreaux1986}. Due to the long coherence time of the nuclear spin and the ability of absolute frequency measurements, comagnetometers based on nuclear magnetic resonances (NMR) are preferred in many precision measurements~\cite{Majumder1990,tullney2013,sachdeva2019}. When alkali atoms are added into the system, the nuclear spin atoms can be both hyperpolarized~\cite{bouchiat1960} and probed~\cite{grover1978} by the polarized alkali atoms, where the nuclear spin signal detected by the in-situ Rb magnetometer is largely amplified due to the Fermi-contact interactions between nuclear spin atoms and alkali atoms~\cite{walker1997}. This hybrid system relaxes the limitations on lasers for optical pumping, speeds up the experiment cycle, and has wide applications in fundamental physics research~\cite{bulatowicz2013,sheng14a,limes18}.

There is also intense interest on using this comagnetometer system as an atomic gyroscope~\cite{donley2009,donley2013}. Among the noble gas elements, Xe has the largest spin-exchange collision rate with the alkali atoms and in turn the highest hyperpolarization rate. Moreover, Xe isotopes experience similar magnetic fields from the polarized alkali atoms~\cite{feng2020}, which largely suppresses systematic effects from the polarized electron spin in this hybrid system. Therefore, the system of Xe isotopes mixed with alkali atoms is most suitable for inertial sensing applications. Pioneered by the Litton Systems Inc.~\cite{grover1979,lam1983} and later Northrop Grumman Corp.~\cite{larsen2014}, the NMR gyroscope has been developed to reach an ARW of 0.005~$^\circ$/h$^{1/2}$, and a bias instability of 0.02~$^\circ$/h with a package size of 10~cm$^3$~\cite{walker16}. Following this breakthrough, there are mainly two research directions in this field. One is to furthur miniaturize the system and migrate multiple similar systems together to realize more functions. The other one is to focus on the limitations of the current system, and work on new detection schemes for potentially better performance~\cite{kover2015,thrasher2019}.

One important goal in developing the system is to realize a highly sensitive Rb magnetometer, because its sensitivity fundamentally limits the signal-to-noise ratio of the comagnetometer. This is challenging to achieve because the presence of numerous Xe gases significantly increases the relaxation rate of Rb atoms due to the collisions between them~\cite{walker1997}. There are both active and passive methods to improve the Rb magnetometer sensitivity. The parametric modulation of the bias field is one of the commonly used active methods~\cite{slocum1973,volk1980,Li2006}. To reach the best sensitivity, the amplitude of the modulation field in this method is required to be comparable with the magnitude of the dc bias field, which potentially affects the comagnetometer stability. On the other hand, the use of multipass cavity is an example of the passive method, which can increase the signal without changing the size or damaging the stability of the system. Due to the simplicity of design and use, Herriott cavities~\cite{Herriott1965,silver2005} are particularly suitable for the applications based on atomic cells, and have been widely applied in atomic magnetometry~\cite{shi2013,sheng2013,cooper2016,cai2020,limes2020,lucivero2021}.

In this paper, we demonstrate a closed-loop $^{129}$Xe-$^{131}$Xe-Rb comagnetometer assisted by a Herriott cavity. In this system, the nuclear spin atoms are continuously driven by oscillating magnetic fields, whose frequencies are kept on resonance by PLLs. This system has been developed to reach an ARW of 0.06$^\circ$/h$^{1/2}$, and a bias instability of 0.2~$^\circ$/h (0.15~$\mu$Hz), according to the standard Allan deviation analysis, with a bandwidth of 1.5~Hz. Moreover, by including an extra closed-loop Rb isotope comagnetometer, this system can be extended to dual simultaneously working comagnetometers sharing the same cell, which also has wide applications in precision measurements. Following this introduction, Sec.~II presents the experiment setup and theoretical background, Sec.~III shows the results and discussions, and Sec.~IV concludes the paper.

\section{Experiment setup and Theoretical background}

\begin{figure}[htp]
\includegraphics[width=3in]{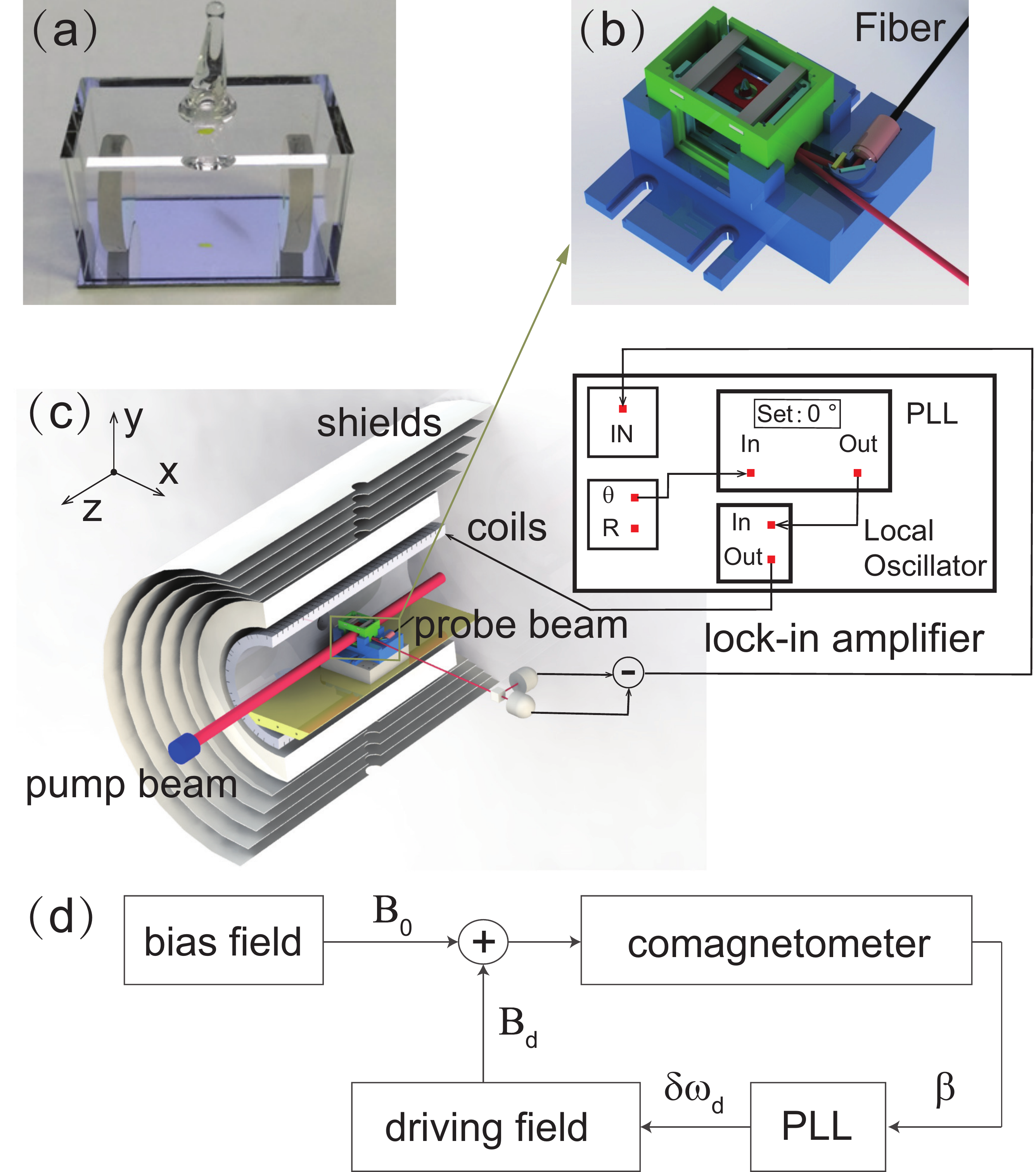}
\caption{\label{fig:setup}(Color online)  Plot(a) shows the atomic cell used in the experiment with a Herriott cavity inside and an inner dimension of 15~mm$\times$15~mm$\times$27~mm, plot(b) shows a 3D printed optical platform for the atomic cell in plot(a), plot(c) shows the experiment setup where we only include a single species operation to simplify the plot, and plot(d) shows the feedback loop.}
\end{figure}

The Herriott cavity used in this paper consists of two cylindrical mirrors with a curvature of 100~mm, a diameter of 12.7~mm, a thickness of 2.5~mm, a relative angle between symmetrical axes of 52$^\circ$, and a separation of 19.3~mm~\cite{silver2005,cai2020}. The cavity mirrors are anodic bonded~\cite{cai2020} to a piece of silicon wafer with a thickness of 0.5 mm. During the bonding process, the mirrors are held at the designed positions on the silicon piece using a ceramic mould, which limits the mirror misalignment within 0.1~mm. A closed cell containing this Herriott cavity is realized by furthur anodic bonding a glass cover to the silicon piece. We then connect the bonded cell to a vacuum string, fill into the cell with 3~torr $^{129}$Xe, 37~torr $^{131}$Xe, 150~torr N$_{2}$, 5~torr H$_{2}$~\cite{kwon81} and Rb atoms of natural abundance, and take off the cell from the vacuum string by a torch. Figure~\ref{fig:setup}(a) shows a picture of the atomic cell with an inner dimension of 15~mm$\times$15~mm$\times$27~mm.

The cell produced using the procedure above is ready to be used together with a 3D printed optical platform as shown in Fig.~\ref{fig:setup}(b). The cell temperature is heated to 110$^\circ$C by running high frequency ($\sim$300~kHz) ac currents through two ceramic heaters placed on the bottom and top of the cell. This arrangement of the heaters makes use of the high thermal conductivity of silicon, helps to heat the cavity mirrors locally, and avoids the coating of Rb atoms on the mirrors~\cite{li2011}. A linearly polarized probe beam is fiber coupled to the optical platform, where it passes through a polarizer, enters the cavity from a 2.5~mm hole in the center of the front mirror, and exits from the same hole after 21 times of reflections.

In the experiment, the optical platform sits in the middle of five-layer mu-metal shields, as shown in Fig.~\ref{fig:setup}(c). A set of solenoid coils and a home-made current supply~\cite{libbrecht1993} are used to generate a bias field of 160~mG in the $z$ direction, and two sets of cosine-theta coils are used for the transverse fields inside the shields. The probe beam on the optical platform, with a detuning of 90 GHz from the Rb D1 transition and a power of 1~mW, is aligned close to the $x$ axis. The Faraday rotation of the transmitted probe beam is analyzed by a polarimeter, consisting of a half-wave plate, a polarization beam splitter and a set of differential photodiode detectors. A circularly polarized pump beam on resonance with the Rb  D1 transition, has a power of 40 mW and an elliptical area of 2 $\times$ 1~cm$^2$, and passes the cell through the $z$ direction. The absorption of the pump beam along its propagation direction leads to a significant Rb polarization gradient in the same direction, which causes an effective magnetic field gradient for the Xe atoms. We add a pair of anti-Helmholtz coils in the $z$ direction to compensate the linear part of this effect. The power of both the pump and probe beams are locked using the methods in Ref.~\cite{cai2020}, and the frequency of the pump beam is locked to a stable single-mode He-Ne laser using a scanning Fabry-Perot cavity~\cite{zhao1998}, which reduces the drift of the pump beam frequency to less than 10~MHz in a day.

\begin{figure}[htb]
\includegraphics[width=3in]{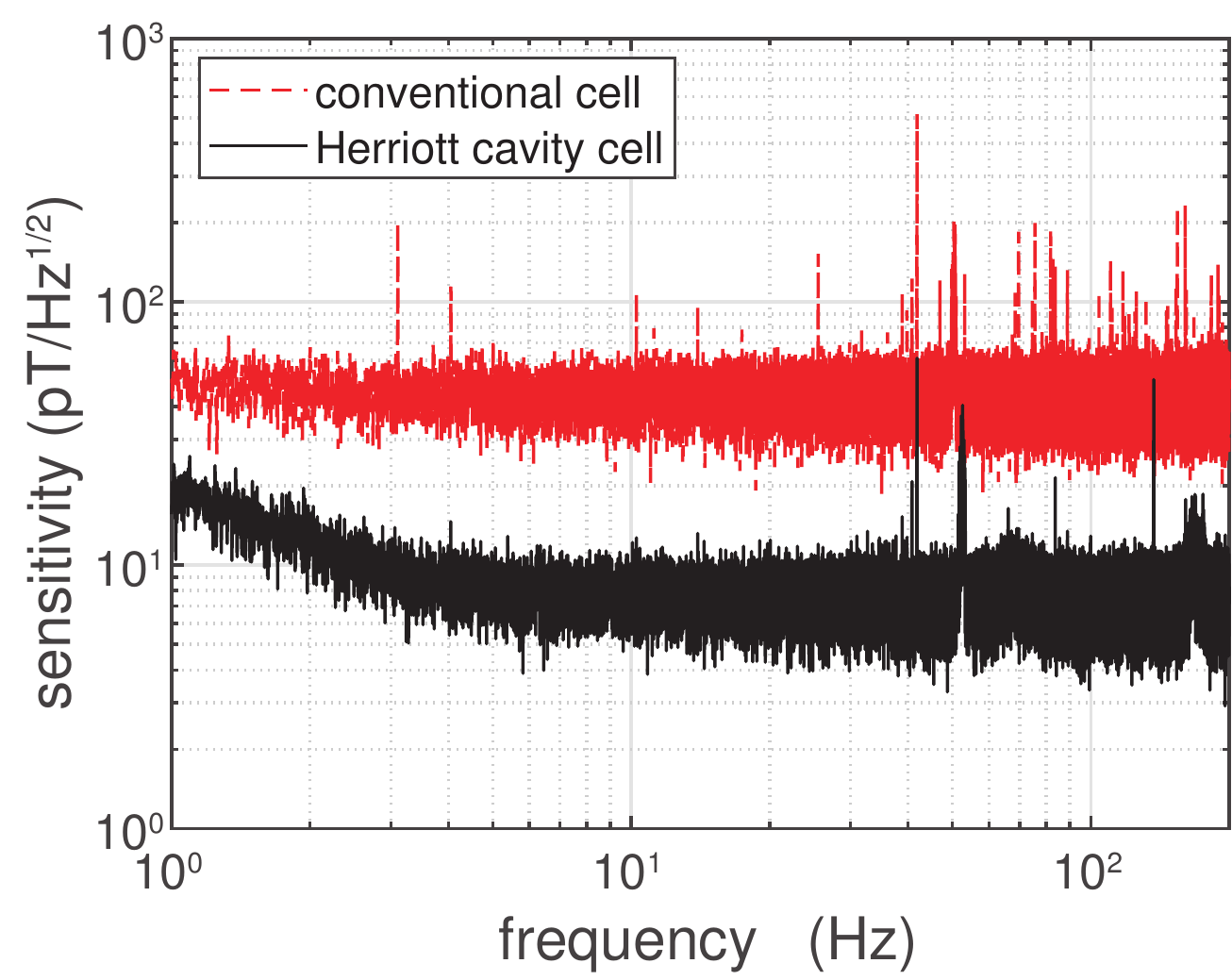}
\caption{\label{fig:Rbmag}(Color online) Comparisons $^{85}$Rb scalar magnetometer sensitivities in a Herriott-cavity-assisted cell (lower solid line) and a conventional vapor cell (upper dash line) under the same experiment conditions.}
\end{figure}

The performance of the Herriott cavity is tested by comparing the Rb magnetometer sensitivities of a Herriott-cavity-assisted cell and a conventional cell, where both cells are prepared with the same experiment conditions except that the conventional cell has an inner dimension of 8~mm$\times$8~mm$\times$8~mm. Fig.~\ref{fig:Rbmag} shows the measurement results of the $^{85}$Rb scalar magnetometer sensitivity using the method in Ref.~\cite{smullin2009}. It demonstrates that the Herriott cavity helps to improve the Rb magnetometer sensitivity by one order of magnitude, which is comparable with the parametric modulation method.

While Xe atoms are pumped by the polarized Rb atoms in the $z$ direction, there are also two oscillating magnetic fields, with frequencies close to the magnetic resonance of Xe isotopes, in the $y$ direction that continuously drive the Xe polarizations away from the $z$ direction. The driving fields are provided by the reference signals of a lock-in amplifier, which also demodulates the comagnetometer signal as shown in Fig.~\ref{fig:setup}(c). The dynamics of the Xe polarization $\mathbf{K}$ are described by the Bloch equation,
\begin{equation}~\label{eq:bloch}
\frac{d\mathbf{K}}{dt}=\mathbf{K}\times(\mathbf{\Omega}_0+\mathbf{\Omega}_d)-\Gamma\mathbf{K}+R_{se}\hat{z}.
\end{equation}
Here, $R_{se}$ corresponds to the rate of spin-exchange collisions with the polarized alkali atoms, and $\mathbf{\Omega}_{0,d}=\gamma\mathbf{B}_{0,d}$, where $\gamma$ is the gyromagnetic ratio, $\mathbf{B}_0$ and $\mathbf{B}_d$ denote the bias field along the $z$ axis and the oscillation field in the $y$ direction, respectively. Following the treatment of Ref.~\cite{walker16}, we introduce two parameters $K_+$ and $\beta$, where $K_+=K_x+iK_y=K_{\bot}e^{-i\phi}$, and $\beta$ is the phase delay between $K_x$ and $\mathbf{B}_d$, so that $\mathbf{B}_d=B_d\cos(\phi-\beta)\hat{y}$. Using the rotating wave approximation~\cite{walker16}, Eq.~\eqref{eq:bloch} becomes:
\begin{eqnarray}
\frac{dK_\bot}{dt}&=&-\Gamma_2K_\bot-\frac{\Omega_dK_z}{2}\cos\beta,\\
\label{eq:phi1}\frac{d\phi}{dt}&=&\Omega_0+\frac{\Omega_dK_z}{2K_\bot}\sin\beta,\\
\frac{dK_z}{dt}&=&\Omega_dK_\bot\cos(\phi-\beta)\cos\phi-\Gamma_1K_z+R_{se}.
\end{eqnarray}
$\Gamma_1$ and $\Gamma_2$ in the equations above denote the longitudinal and transverse depolarization rate of the nuclear spin atom, respectively. These equations are slightly different from Eqs. (11)-(13) in Ref.~\cite{walker16} because the driving fields are applied in different directions. The steady state solutions of these equations are:
\begin{eqnarray}
\label{eq:beta}\beta&=&\tan^{-1}\left(\frac{\Omega_0-\omega_d}{\Gamma_2}\right),\\
K_z&=&\frac{R_{se}}{\Gamma_1+\frac{\Omega_d^2}{4\Gamma_2}\cos^2\beta},\\
\label{eq:Kamp}K_\bot&=&-\frac{\Omega_dK_z}{2\Gamma_2}\cos\beta,
\end{eqnarray}
where $\omega_d$ is the frequency of the driving field.

\section{Results and Discussions}
\begin{figure}[htb]
\includegraphics[width=3in]{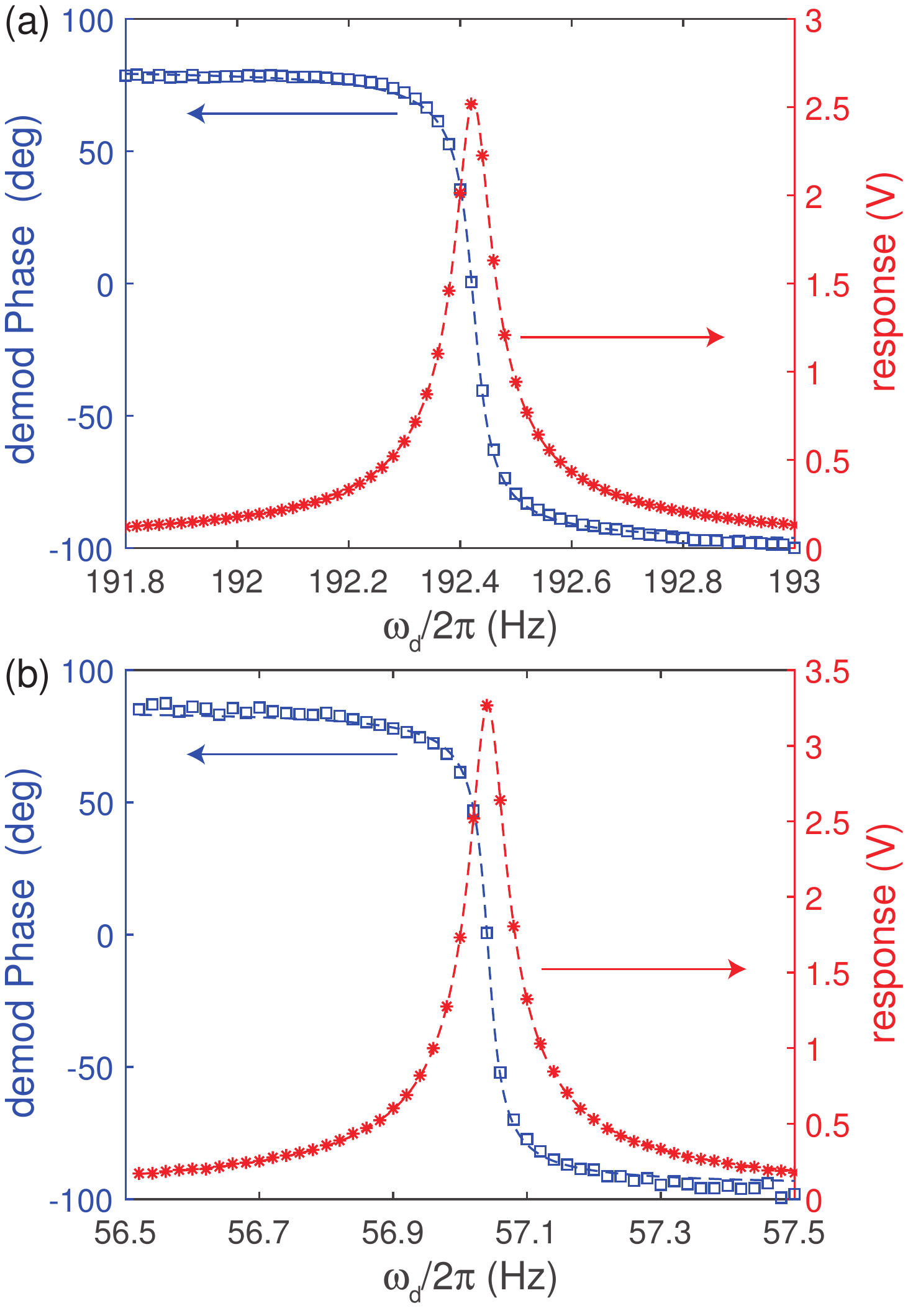}
\caption{\label{fig:phasetheta}(Color online) Plot(a) and (b) show the demodulated amplitude and phase of $^{129}$Xe and $^{131}$Xe signals as a function of $\omega_d$. The points are experiment data, and the dash lines are fitting results using Eqs.~\eqref{eq:beta} and~\eqref{eq:Kamp}. }
\end{figure}

From Eqs.~\eqref{eq:beta}-\eqref{eq:Kamp}, we can conclude that the comagnetometer signal, proportional to $K_x$, shows the maximum demodulated amplitude and the largest phase response to the external field changes  when $\omega_d$ is on resonance with the nuclear spin Larmor frequency. Figure~\ref{fig:phasetheta}(a) shows the experiment results of the demodulated $^{129}$Xe signals as a function of the driving field frequency in the open-loop operation, with a driving field amplitude $B_d=1$~nT. The experiment data shows a good agreement with the fitting results using Eqs.~\eqref{eq:beta} and~\eqref{eq:Kamp}, from which we extract $\Gamma_2/2\pi$=24~mHz. Similarly, Fig.~\ref{fig:phasetheta}(b) demonstrates the demodulated results of $^{131}$Xe signals with a driving field amplitude $B_d=4$~nT, which shows a transverse depolarization rate of $\Gamma_2/2\pi$=16~mHz.

It is clear from Fig.~\ref{fig:phasetheta} that, for each Xe isotope, the dependence of its demodulated phase on the driving field detuning can be used for the magnetic field sensing. A closed-loop operation is preferred because it is more robust and convenient to use.  As shown in Fig.~\ref{fig:setup}(d), we can reach a closed-loop operation by sending the phase results to a PLL, where the set point of the phase is zero and the feedback signal is used to control the driving field frequency $\omega_d$. Therefore, any magnetic field change would lead to a corresponding change in $\omega_d$ , so that the nuclear spin is always driven by resonant fields, and we can record the data of $\omega_d$ to track the magnetic field information. This operation is also valid  for the alkali atom magnetometer.

\begin{figure}[hpt]
\includegraphics[width=3in]{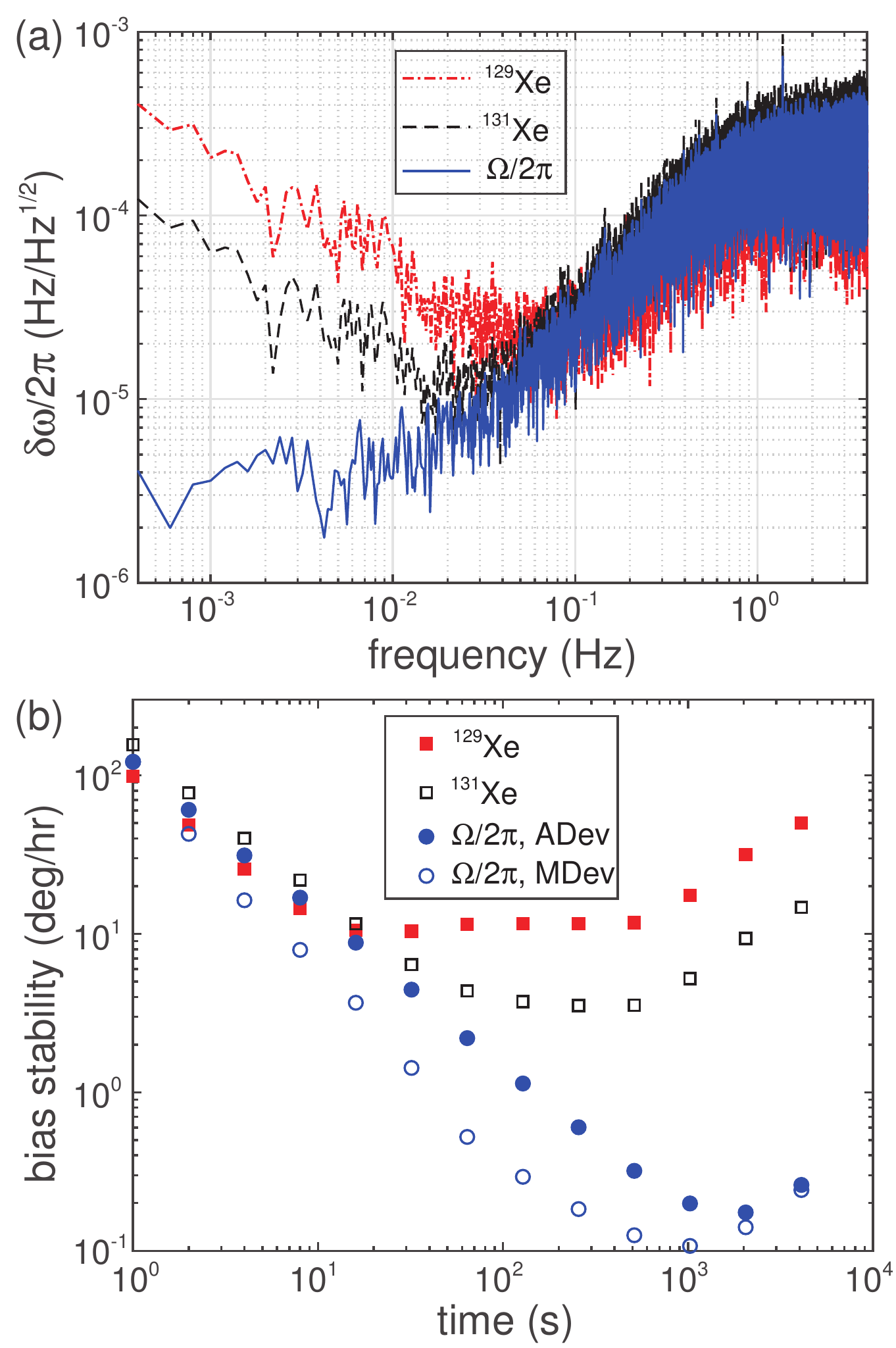}
\caption{\label{fig:noise}(Color online) Plot(a) shows amplitude spectral densities of the recorded frequencies of $^{129}$Xe (upper dash dotted curve), $^{131}$Xe (middle dash curve) and the rotation (lower solid curve). Plot(b) shows the standard Allan deviation (ADev) and modified Allan deviation (MDev)  results of the same data in plot(a). }
\end{figure}

A closed-loop Xe isotope comagnetometer contains two closed-loop Xe mangetometers, and its gyroscopic response of the rotation along the bias field can be extracted from the combination of the recorded frequencies of both Xe isotopes:
\begin{equation}
\Omega=\frac{\omega_a-R\omega_b}{R-1},
\end{equation}
where $a$ and $b$ denote $^{129}$Xe and $^{131}$Xe, respectively, and $R\approx\gamma_a/\gamma_b=-3.373\cdots$. Figure~\ref{fig:noise}(a) shows the amplitude spectral densities of $\omega_a$, $\omega_b$ and $\Omega$. For each Xe isotope, its precession frequency noise is dominated by the white phase noise in the frequency range of 0.1~Hz - 1~Hz, shows a signature of white frequency noise in the frequency range of 0.01 Hz - 0.1 Hz, and is mainly the random walk frequency noise in the frequency range below 0.01 Hz. Most contributions to noise in the measurement of $\omega_{a,b}$ in the low frequency domain are caused by the fluctuations or drift of the bias field and the effective field from the polarized Rb atoms, which are common-mode for both Xe isotopes and can be cancelled when extracting $\Omega$.

The noise of $\Omega$ is mainly the white and flicker phase noise in the frequency range of 0.01 Hz - 0.1 Hz, and dominated by the white frequency noise in the frequency range of 0.001 Hz - 0.01 Hz. This leads to an integration time of 2000 s for the rotation frequency data in the time domain to reach a bias instability of 0.2~$^\circ$/h (0.15~$\mu$Hz), as shown by the standard Allan deviation analysis in Fig.~\ref{fig:noise}(b). In the modified Allan deviation plot, the bias instability is a factor of two smaller~\cite{rubiola2009}. The ARW of the comagnetometer is the slope of the $t^{-1/2}$ part in the standard Allan deviation plot, and it is equal to $h_0/\sqrt{2}$~\cite{rubiola2009}, where $h_0$ is the white frequency noise level of $\Omega$. Using the data in Fig.~\ref{fig:noise}(a), we get $h_0=4.2$~$\mu$Hz/Hz$^{1/2}$ and the ARW as 0.06$^\circ$/h$^{1/2}$ .

\begin{figure}[htb]
\includegraphics[width=3in]{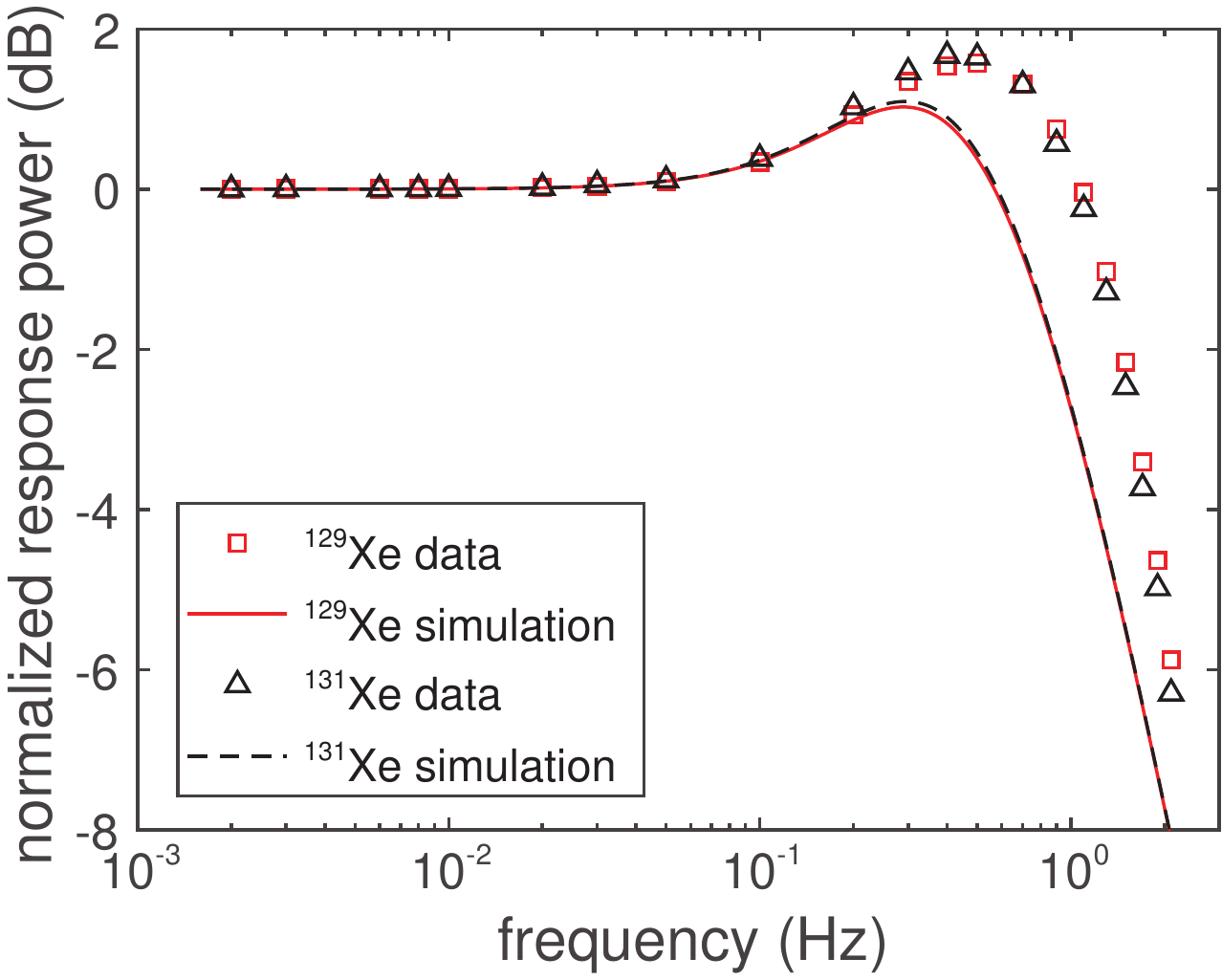}
\caption{\label{fig:band}(Color online) The normalized power of the closed-loop Xe magnetometers response to the bias field modulations with a modulation field amplitude $B_m=1$~nT. The points are experiment data, and the lines are simulation results using Eq.~\eqref{eq:Gs} with experiment conditions.}
\end{figure}

Another advantage of the closed-loop operation is that its bandwidth is not limited by the open-loop line width  anymore. By adding a modulation field $B_m\sin(\omega_m{t})$ to the bias field and recording the power of the corresponding peak at $\omega_m$ in the amplitude spectrum of the comagnetometer output,  we show the bandwidth measurement results in Fig.~\ref{fig:band}. The results demonstrate a closed-loop bandwidth of 1.5~Hz for both Xe isotopes, about two orders of magnitude increase compared with the open-loop results in Fig.~\ref{fig:phasetheta}. To better understand the system performance, we also simulate the closed-loop response of the comagnetometer using the method in Ref.~\cite{zhang2020}. For a small perturbation on the bias field, the transfer function of the open-loop comagnetometer can be extracted from Eq.~\eqref{eq:phi1} as
\begin{equation}
H(s)=\left(\frac{\delta\beta}{\delta\Omega_0}\right)_o=\frac{1}{s+\Gamma_2}.
\end{equation}
The other part of the loop in Fig.~\ref{fig:setup}(d) is the PLL, whose transfer function is:
\begin{equation}
F(s)=\frac{\delta{\omega}_d}{\delta{\beta}}=c_P+\frac{c_I}{s},
\end{equation}
where $c_P$ and $c_I$ are the coefficients of the proportion and integration parts of the feedback loop, and the derivative part is neglected because it is not used in the experiment. Then, the closed-loop response can be expressed as
\begin{eqnarray}
\label{eq:Gs}G(s)=\left(\frac{\delta\omega_d}{\delta\Omega_0}\right)_c=\frac{H(s)F(s)}{1+H(s)F(s)}\nonumber\\
=\frac{c_Ps+c_I}{s^2+(c_p+\Gamma_2)s+c_I}.
\end{eqnarray}
With the experiment parameters of $c_P=5.4$~s$^{-1}$ and $c_I=7.2$~s$^{-2}$, the simulation results using Eq.~\eqref{eq:Gs} are also plotted in Fig.~\ref{fig:band}. It shows a good agreement with the experiment data in the frequency domain below 0.2~Hz, and the deviations at higher frequencies are still under investigation.

\begin{figure}[htb]
\includegraphics[width=3in]{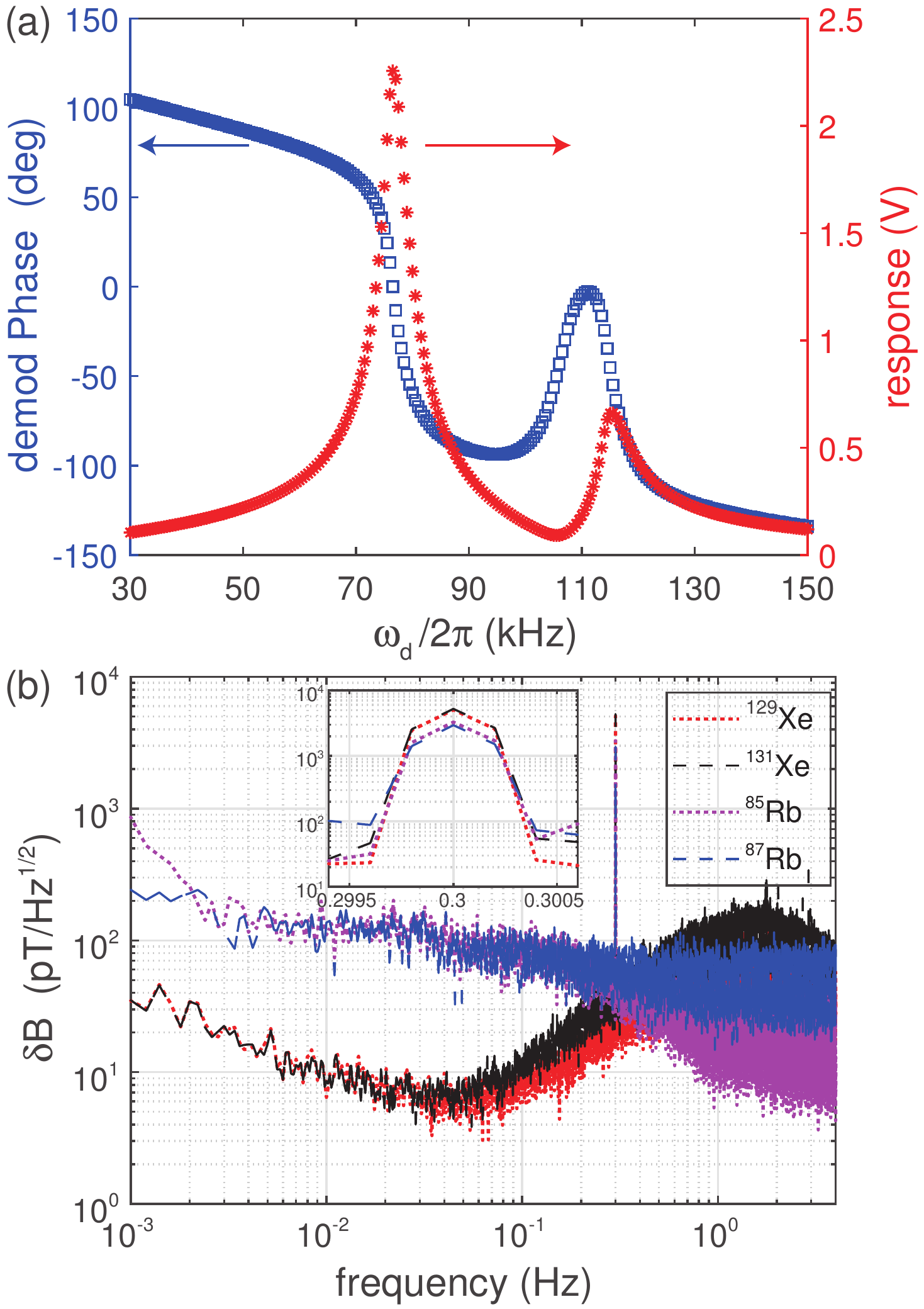}
\caption{\label{fig:Rb}(Color online) Plot(a) shows the open-loop behavior of Rb atoms, and plot(b) shows the amplitude spectrum of the closed-loop Xe (lower curves in the main plot and upper curves in the inset) and Rb magnetometers in presence of a bias field modulation with an amplitude of 0.1~nT and a frequency of 0.3~Hz.}
\end{figure}

The system studied in this paper also provides a flexible platform to combine multiple atomic oscillators. Since there are two Rb isotopes in the cell, we can adapt the methods used in the closed-loop Xe isotope comagnetometer, and develop a similar Rb isotope comagnetometer. Figure~\ref{fig:Rb}(a) shows the demodulated phase and amplitude for Rb isotopes as a function of the driving field frequencies. When both the Xe isotope comagnetometer and the Rb isotope comagnetometer are working in the closed-loop mode, we achieve a system of dual simultaneously working comagnetometers, sharing the same Herriott-cavity-assisted vapor cell and having the same sensitive direction of detecting rotations.

\begin{table}[hptb]
  \begin{center}
    \caption{Experiment parameters of different closed-loop atomic magnetometers.}
    \label{tab:tpara}
    \begin{tabular}{|c|c|c|c|c|} 
    \hline
    &&&&\\
      &$^{129}$Xe &$^{131}$Xe &$^{85}$Rb&$^{87}$Rb\\
      \hline
      $c_P$ (s$^{-1}$)&5.4&5.4&7.2$\times10^4$&7.2$\times10^4$\\
      \hline
      $c_I$ (s$^{-2}$)&7.2&7.2&7.2$\times10^3$&7.2$\times10^3$\\
      \hline
      $\Gamma_2$ (s$^{-1}$)&0.15&0.1&1.6$\times10^4$&1.8$\times10^4$\\
      \hline
    \end{tabular}
  \end{center}
\end{table}

The experimental parameters of closed-loop operations are very different for Xe and Rb atoms as shown in Tab.~\ref{tab:tpara}, and it is interesting to compare their responses to external fields in these conditions. For low frequencies, we can neglect the $s^2$ term in the denominator of Eq.~\eqref{eq:Gs} so that,
\begin{equation}
\label{eq:h6}
G(s)\approx\frac{c_P s+c_I}{(c_P+\Gamma_2) s+c_I}.
\end{equation}
It is clear from this equation that the amplitude of the response decreases when $\Gamma_2$ increases, and it is equal to 1 only in the low relaxation rate limit. Figure~\ref{fig:Rb}(b) demonstrates the experimental data of the amplitude spectrum of these closed-loop magnetometers subject to a weak perturbation in the bias field at the frequency of 0.3~Hz. The plot shows that, while the response amplitudes are very similar for comparisons between Xe isotopes or Rb isotopes, the response amplitude of the Rb atom is 35\% smaller than that of the Xe atom. In the simulation, using Eq.~\eqref{eq:Gs} and the experiment conditions, we get $|G(s)|_{Xe}\approx1$ and $|G(s)|_{Rb}\approx0.8$ at $\omega=2\pi\times0.3$~Hz.

\section{Conclusion}
In conclusion, we have realized a Herriott-cavity-assisted closed-loop  Xe isotope comagnetometer, where we use the multipass cavity, instead of the commonly used parametric modulation method, to improve the Rb magnetometer sensitivity. This system has demonstrated an ARW of 0.06~$^\circ$/h$^{1/2}$, and a bias instability of 0.2~$^\circ$/h (0.15~$\mu$Hz) with a bandwidth of 1.5~Hz.  We are currently working on a more compact system with a smaller Herriott-cavity vapor cell, which aims to reduce the effects of magnetic field gradients from the polarized Rb atoms and furthur improve the system performance. We are also updating the experiment hardware by adding a rotation table to the testing platform, so that we can calibrate the rotation response of the comagnetometer and measure the effects of the earth rotation.

We also develop a system of dual simultaneously working comagnetometers in this work, by adding a closed-loop Rb isotope comagnetometer. This extended system has potential applications in precision measurements, such as the advanced GNOME experiment~\cite{pospelov2013,pustelny2013,hector2020,afach2021}, where we can simultaneously and independently measure the anomalous coupling of the axion-like particles with proton spin and neutron spin.

\section*{Acknowledgements}
This work was partially carried out at the USTC Center for Micro and Nanoscale Research and Fabrication. We thank X.-D. Zhang and Dr. N. Zhao for helpful discussions, and S.-B. Zhang for assistances on the experiment. This work was supported by Natural Science Foundation of China (Grant No. 11974329).


\begin{thebibliography}{41}%
\makeatletter
\providecommand \@ifxundefined [1]{%
 \@ifx{#1\undefined}
}%
\providecommand \@ifnum [1]{%
 \ifnum #1\expandafter \@firstoftwo
 \else \expandafter \@secondoftwo
 \fi
}%
\providecommand \@ifx [1]{%
 \ifx #1\expandafter \@firstoftwo
 \else \expandafter \@secondoftwo
 \fi
}%
\providecommand \natexlab [1]{#1}%
\providecommand \enquote  [1]{``#1''}%
\providecommand \bibnamefont  [1]{#1}%
\providecommand \bibfnamefont [1]{#1}%
\providecommand \citenamefont [1]{#1}%
\providecommand \href@noop [0]{\@secondoftwo}%
\providecommand \href [0]{\begingroup \@sanitize@url \@href}%
\providecommand \@href[1]{\@@startlink{#1}\@@href}%
\providecommand \@@href[1]{\endgroup#1\@@endlink}%
\providecommand \@sanitize@url [0]{\catcode `\\12\catcode `\$12\catcode
  `\&12\catcode `\#12\catcode `\^12\catcode `\_12\catcode `\%12\relax}%
\providecommand \@@startlink[1]{}%
\providecommand \@@endlink[0]{}%
\providecommand \url  [0]{\begingroup\@sanitize@url \@url }%
\providecommand \@url [1]{\endgroup\@href {#1}{\urlprefix }}%
\providecommand \urlprefix  [0]{URL }%
\providecommand \Eprint [0]{\href }%
\providecommand \doibase [0]{http://dx.doi.org/}%
\providecommand \selectlanguage [0]{\@gobble}%
\providecommand \bibinfo  [0]{\@secondoftwo}%
\providecommand \bibfield  [0]{\@secondoftwo}%
\providecommand \translation [1]{[#1]}%
\providecommand \BibitemOpen [0]{}%
\providecommand \bibitemStop [0]{}%
\providecommand \bibitemNoStop [0]{.\EOS\space}%
\providecommand \EOS [0]{\spacefactor3000\relax}%
\providecommand \BibitemShut  [1]{\csname bibitem#1\endcsname}%
\let\auto@bib@innerbib\@empty
\bibitem [{\citenamefont {Lamoreaux}\ \emph {et~al.}(1986)\citenamefont
  {Lamoreaux}, \citenamefont {Jacobs}, \citenamefont {Heckel}, \citenamefont
  {Raab},\ and\ \citenamefont {Fortson}}]{lamoreaux1986}%
  \BibitemOpen
  \bibfield  {author} {\bibinfo {author} {\bibfnamefont {S.~K.}\ \bibnamefont
  {Lamoreaux}}, \bibinfo {author} {\bibfnamefont {J.~P.}\ \bibnamefont
  {Jacobs}}, \bibinfo {author} {\bibfnamefont {B.~R.}\ \bibnamefont {Heckel}},
  \bibinfo {author} {\bibfnamefont {F.~J.}\ \bibnamefont {Raab}}, \ and\
  \bibinfo {author} {\bibfnamefont {E.~N.}\ \bibnamefont {Fortson}},\ }\href
  {\doibase 10.1103/PhysRevLett.57.3125} {\bibfield  {journal} {\bibinfo
  {journal} {Phys. Rev. Lett.}\ }\textbf {\bibinfo {volume} {57}},\ \bibinfo
  {pages} {3125} (\bibinfo {year} {1986})}\BibitemShut {NoStop}%
\bibitem [{\citenamefont {Majumder}\ \emph {et~al.}(1990)\citenamefont
  {Majumder}, \citenamefont {Venema}, \citenamefont {Lamoreaux}, \citenamefont
  {Heckel},\ and\ \citenamefont {Fortson}}]{Majumder1990}%
  \BibitemOpen
  \bibfield  {author} {\bibinfo {author} {\bibfnamefont {P.~K.}\ \bibnamefont
  {Majumder}}, \bibinfo {author} {\bibfnamefont {B.~J.}\ \bibnamefont
  {Venema}}, \bibinfo {author} {\bibfnamefont {S.~K.}\ \bibnamefont
  {Lamoreaux}}, \bibinfo {author} {\bibfnamefont {B.~R.}\ \bibnamefont
  {Heckel}}, \ and\ \bibinfo {author} {\bibfnamefont {E.~N.}\ \bibnamefont
  {Fortson}},\ }\href {\doibase 10.1103/PhysRevLett.65.2931} {\bibfield
  {journal} {\bibinfo  {journal} {Phys. Rev. Lett.}\ }\textbf {\bibinfo
  {volume} {65}},\ \bibinfo {pages} {2931} (\bibinfo {year}
  {1990})}\BibitemShut {NoStop}%
\bibitem [{\citenamefont {Tullney}\ \emph {et~al.}(2013)\citenamefont
  {Tullney}, \citenamefont {Allmendinger}, \citenamefont {Burghoff},
  \citenamefont {Heil}, \citenamefont {Karpuk}, \citenamefont {Kilian},
  \citenamefont {Knappe-Gr\"uneberg}, \citenamefont {M\"uller}, \citenamefont
  {Schmidt}, \citenamefont {Schnabel}, \citenamefont {Seifert}, \citenamefont
  {Sobolev},\ and\ \citenamefont {Trahms}}]{tullney2013}%
  \BibitemOpen
  \bibfield  {author} {\bibinfo {author} {\bibfnamefont {K.}~\bibnamefont
  {Tullney}}, \bibinfo {author} {\bibfnamefont {F.}~\bibnamefont
  {Allmendinger}}, \bibinfo {author} {\bibfnamefont {M.}~\bibnamefont
  {Burghoff}}, \bibinfo {author} {\bibfnamefont {W.}~\bibnamefont {Heil}},
  \bibinfo {author} {\bibfnamefont {S.}~\bibnamefont {Karpuk}}, \bibinfo
  {author} {\bibfnamefont {W.}~\bibnamefont {Kilian}}, \bibinfo {author}
  {\bibfnamefont {S.}~\bibnamefont {Knappe-Gr\"uneberg}}, \bibinfo {author}
  {\bibfnamefont {W.}~\bibnamefont {M\"uller}}, \bibinfo {author}
  {\bibfnamefont {U.}~\bibnamefont {Schmidt}}, \bibinfo {author} {\bibfnamefont
  {A.}~\bibnamefont {Schnabel}}, \bibinfo {author} {\bibfnamefont
  {F.}~\bibnamefont {Seifert}}, \bibinfo {author} {\bibfnamefont
  {Y.}~\bibnamefont {Sobolev}}, \ and\ \bibinfo {author} {\bibfnamefont
  {L.}~\bibnamefont {Trahms}},\ }\href {\doibase
  10.1103/PhysRevLett.111.100801} {\bibfield  {journal} {\bibinfo  {journal}
  {Phys. Rev. Lett.}\ }\textbf {\bibinfo {volume} {111}},\ \bibinfo {pages}
  {100801} (\bibinfo {year} {2013})}\BibitemShut {NoStop}%
\bibitem [{\citenamefont {Sachdeva}\ \emph {et~al.}(2019)\citenamefont
  {Sachdeva}, \citenamefont {Fan}, \citenamefont {Babcock}, \citenamefont
  {Burghoff}, \citenamefont {Chupp}, \citenamefont {Degenkolb}, \citenamefont
  {Fierlinger}, \citenamefont {Haude}, \citenamefont {Kraegeloh}, \citenamefont
  {Kilian}, \citenamefont {Knappe-Gr\"uneberg}, \citenamefont {Kuchler},
  \citenamefont {Liu}, \citenamefont {Marino}, \citenamefont {Meinel},
  \citenamefont {Rolfs}, \citenamefont {Salhi}, \citenamefont {Schnabel},
  \citenamefont {Singh}, \citenamefont {Stuiber}, \citenamefont {Terrano},
  \citenamefont {Trahms},\ and\ \citenamefont {Voigt}}]{sachdeva2019}%
  \BibitemOpen
  \bibfield  {author} {\bibinfo {author} {\bibfnamefont {N.}~\bibnamefont
  {Sachdeva}}, \bibinfo {author} {\bibfnamefont {I.}~\bibnamefont {Fan}},
  \bibinfo {author} {\bibfnamefont {E.}~\bibnamefont {Babcock}}, \bibinfo
  {author} {\bibfnamefont {M.}~\bibnamefont {Burghoff}}, \bibinfo {author}
  {\bibfnamefont {T.~E.}\ \bibnamefont {Chupp}}, \bibinfo {author}
  {\bibfnamefont {S.}~\bibnamefont {Degenkolb}}, \bibinfo {author}
  {\bibfnamefont {P.}~\bibnamefont {Fierlinger}}, \bibinfo {author}
  {\bibfnamefont {S.}~\bibnamefont {Haude}}, \bibinfo {author} {\bibfnamefont
  {E.}~\bibnamefont {Kraegeloh}}, \bibinfo {author} {\bibfnamefont
  {W.}~\bibnamefont {Kilian}}, \bibinfo {author} {\bibfnamefont
  {S.}~\bibnamefont {Knappe-Gr\"uneberg}}, \bibinfo {author} {\bibfnamefont
  {F.}~\bibnamefont {Kuchler}}, \bibinfo {author} {\bibfnamefont
  {T.}~\bibnamefont {Liu}}, \bibinfo {author} {\bibfnamefont {M.}~\bibnamefont
  {Marino}}, \bibinfo {author} {\bibfnamefont {J.}~\bibnamefont {Meinel}},
  \bibinfo {author} {\bibfnamefont {K.}~\bibnamefont {Rolfs}}, \bibinfo
  {author} {\bibfnamefont {Z.}~\bibnamefont {Salhi}}, \bibinfo {author}
  {\bibfnamefont {A.}~\bibnamefont {Schnabel}}, \bibinfo {author}
  {\bibfnamefont {J.~T.}\ \bibnamefont {Singh}}, \bibinfo {author}
  {\bibfnamefont {S.}~\bibnamefont {Stuiber}}, \bibinfo {author} {\bibfnamefont
  {W.~A.}\ \bibnamefont {Terrano}}, \bibinfo {author} {\bibfnamefont
  {L.}~\bibnamefont {Trahms}}, \ and\ \bibinfo {author} {\bibfnamefont
  {J.}~\bibnamefont {Voigt}},\ }\href {\doibase 10.1103/PhysRevLett.123.143003}
  {\bibfield  {journal} {\bibinfo  {journal} {Phys. Rev. Lett.}\ }\textbf
  {\bibinfo {volume} {123}},\ \bibinfo {pages} {143003} (\bibinfo {year}
  {2019})}\BibitemShut {NoStop}%
\bibitem [{\citenamefont {Bouchiat}\ \emph {et~al.}(1960)\citenamefont
  {Bouchiat}, \citenamefont {Carver},\ and\ \citenamefont
  {Varnum}}]{bouchiat1960}%
  \BibitemOpen
  \bibfield  {author} {\bibinfo {author} {\bibfnamefont {M.~A.}\ \bibnamefont
  {Bouchiat}}, \bibinfo {author} {\bibfnamefont {T.~R.}\ \bibnamefont
  {Carver}}, \ and\ \bibinfo {author} {\bibfnamefont {C.~M.}\ \bibnamefont
  {Varnum}},\ }\href {\doibase DOI 10.1103/PhysRevLett.5.373} {\bibfield
  {journal} {\bibinfo  {journal} {Phys. Rev. Lett.}\ }\textbf {\bibinfo
  {volume} {5}},\ \bibinfo {pages} {373} (\bibinfo {year} {1960})}\BibitemShut
  {NoStop}%
\bibitem [{\citenamefont {Grover}(1978)}]{grover1978}%
  \BibitemOpen
  \bibfield  {author} {\bibinfo {author} {\bibfnamefont {B.~C.}\ \bibnamefont
  {Grover}},\ }\href {\doibase DOI 10.1103/PhysRevLett.40.391} {\bibfield
  {journal} {\bibinfo  {journal} {Phys. Rev. Lett.}\ }\textbf {\bibinfo
  {volume} {40}},\ \bibinfo {pages} {391} (\bibinfo {year} {1978})}\BibitemShut
  {NoStop}%
\bibitem [{\citenamefont {Walker}\ and\ \citenamefont
  {Happer}(1997)}]{walker1997}%
  \BibitemOpen
  \bibfield  {author} {\bibinfo {author} {\bibfnamefont {T.~G.}\ \bibnamefont
  {Walker}}\ and\ \bibinfo {author} {\bibfnamefont {W.}~\bibnamefont
  {Happer}},\ }\href {\doibase DOI 10.1103/RevModPhys.69.629} {\bibfield
  {journal} {\bibinfo  {journal} {Rev. Mod. Phys.}\ }\textbf {\bibinfo {volume}
  {69}},\ \bibinfo {pages} {629} (\bibinfo {year} {1997})}\BibitemShut
  {NoStop}%
\bibitem [{\citenamefont {Bulatowicz}\ \emph {et~al.}(2013)\citenamefont
  {Bulatowicz}, \citenamefont {Griffith}, \citenamefont {Larsen}, \citenamefont
  {Mirijanian}, \citenamefont {Fu}, \citenamefont {Smith}, \citenamefont
  {Snow}, \citenamefont {Yan},\ and\ \citenamefont {Walker}}]{bulatowicz2013}%
  \BibitemOpen
  \bibfield  {author} {\bibinfo {author} {\bibfnamefont {M.}~\bibnamefont
  {Bulatowicz}}, \bibinfo {author} {\bibfnamefont {R.}~\bibnamefont
  {Griffith}}, \bibinfo {author} {\bibfnamefont {M.}~\bibnamefont {Larsen}},
  \bibinfo {author} {\bibfnamefont {J.}~\bibnamefont {Mirijanian}}, \bibinfo
  {author} {\bibfnamefont {C.~B.}\ \bibnamefont {Fu}}, \bibinfo {author}
  {\bibfnamefont {E.}~\bibnamefont {Smith}}, \bibinfo {author} {\bibfnamefont
  {W.~M.}\ \bibnamefont {Snow}}, \bibinfo {author} {\bibfnamefont
  {H.}~\bibnamefont {Yan}}, \ and\ \bibinfo {author} {\bibfnamefont {T.~G.}\
  \bibnamefont {Walker}},\ }\href {\doibase 10.1103/PhysRevLett.111.102001}
  {\bibfield  {journal} {\bibinfo  {journal} {Phys. Rev. Lett.}\ }\textbf
  {\bibinfo {volume} {111}},\ \bibinfo {pages} {102001} (\bibinfo {year}
  {2013})}\BibitemShut {NoStop}%
\bibitem [{\citenamefont {Sheng}\ \emph {et~al.}(2014)\citenamefont {Sheng},
  \citenamefont {Kabcenell},\ and\ \citenamefont {Romalis}}]{sheng14a}%
  \BibitemOpen
  \bibfield  {author} {\bibinfo {author} {\bibfnamefont {D.}~\bibnamefont
  {Sheng}}, \bibinfo {author} {\bibfnamefont {A.}~\bibnamefont {Kabcenell}}, \
  and\ \bibinfo {author} {\bibfnamefont {M.~V.}\ \bibnamefont {Romalis}},\
  }\href {\doibase 10.1103/PhysRevLett.113.163002} {\bibfield  {journal}
  {\bibinfo  {journal} {Phys. Rev. Lett.}\ }\textbf {\bibinfo {volume} {113}},\
  \bibinfo {pages} {163002} (\bibinfo {year} {2014})}\BibitemShut {NoStop}%
\bibitem [{\citenamefont {Limes}\ \emph {et~al.}(2018)\citenamefont {Limes},
  \citenamefont {Sheng},\ and\ \citenamefont {Romalis}}]{limes18}%
  \BibitemOpen
  \bibfield  {author} {\bibinfo {author} {\bibfnamefont {M.~E.}\ \bibnamefont
  {Limes}}, \bibinfo {author} {\bibfnamefont {D.}~\bibnamefont {Sheng}}, \ and\
  \bibinfo {author} {\bibfnamefont {M.~V.}\ \bibnamefont {Romalis}},\ }\href
  {\doibase 10.1103/PhysRevLett.120.033401} {\bibfield  {journal} {\bibinfo
  {journal} {Phys. Rev. Lett.}\ }\textbf {\bibinfo {volume} {120}},\ \bibinfo
  {pages} {033401} (\bibinfo {year} {2018})}\BibitemShut {NoStop}%
\bibitem [{\citenamefont {Donley}\ \emph {et~al.}(2009)\citenamefont {Donley},
  \citenamefont {Long}, \citenamefont {Liebisch}, \citenamefont {Hodby},
  \citenamefont {Fisher},\ and\ \citenamefont {Kitching}}]{donley2009}%
  \BibitemOpen
  \bibfield  {author} {\bibinfo {author} {\bibfnamefont {E.~A.}\ \bibnamefont
  {Donley}}, \bibinfo {author} {\bibfnamefont {J.~L.}\ \bibnamefont {Long}},
  \bibinfo {author} {\bibfnamefont {T.~C.}\ \bibnamefont {Liebisch}}, \bibinfo
  {author} {\bibfnamefont {E.~R.}\ \bibnamefont {Hodby}}, \bibinfo {author}
  {\bibfnamefont {T.~A.}\ \bibnamefont {Fisher}}, \ and\ \bibinfo {author}
  {\bibfnamefont {J.}~\bibnamefont {Kitching}},\ }\href {\doibase
  10.1103/PhysRevA.79.013420} {\bibfield  {journal} {\bibinfo  {journal} {Phys.
  Rev. A}\ }\textbf {\bibinfo {volume} {79}},\ \bibinfo {pages} {013420}
  (\bibinfo {year} {2009})}\BibitemShut {NoStop}%
\bibitem [{\citenamefont {Donley}\ and\ \citenamefont
  {Kitching}(2013)}]{donley2013}%
  \BibitemOpen
  \bibfield  {author} {\bibinfo {author} {\bibfnamefont {E.}~\bibnamefont
  {Donley}}\ and\ \bibinfo {author} {\bibfnamefont {J.}~\bibnamefont
  {Kitching}},\ }in\ \href@noop {} {\emph {\bibinfo {booktitle} {Optical
  Magnetometry}}}\ (\bibinfo  {publisher} {Cambridge University Press, New
  York},\ \bibinfo {year} {2013})\ p.\ \bibinfo {pages} {369}\BibitemShut
  {NoStop}%
\bibitem [{\citenamefont {Feng}\ \emph {et~al.}(2020)\citenamefont {Feng},
  \citenamefont {Zhang}, \citenamefont {Lu},\ and\ \citenamefont
  {Sheng}}]{feng2020}%
  \BibitemOpen
  \bibfield  {author} {\bibinfo {author} {\bibfnamefont {Y.-K.}\ \bibnamefont
  {Feng}}, \bibinfo {author} {\bibfnamefont {S.-B.}\ \bibnamefont {Zhang}},
  \bibinfo {author} {\bibfnamefont {Z.-T.}\ \bibnamefont {Lu}}, \ and\ \bibinfo
  {author} {\bibfnamefont {D.}~\bibnamefont {Sheng}},\ }\href {\doibase
  10.1103/PhysRevA.102.043109} {\bibfield  {journal} {\bibinfo  {journal}
  {Phys. Rev. A}\ }\textbf {\bibinfo {volume} {102}},\ \bibinfo {pages}
  {043109} (\bibinfo {year} {2020})}\BibitemShut {NoStop}%
\bibitem [{\citenamefont {Grover}\ \emph {et~al.}(1979)\citenamefont {Grover},
  \citenamefont {Kanegsberg}, \citenamefont {Mark},\ and\ \citenamefont
  {Meyer}}]{grover1979}%
  \BibitemOpen
  \bibfield  {author} {\bibinfo {author} {\bibfnamefont {B.~C.}\ \bibnamefont
  {Grover}}, \bibinfo {author} {\bibfnamefont {E.}~\bibnamefont {Kanegsberg}},
  \bibinfo {author} {\bibfnamefont {J.~G.}\ \bibnamefont {Mark}}, \ and\
  \bibinfo {author} {\bibfnamefont {R.~L.}\ \bibnamefont {Meyer}},\ }\href@noop
  {} {}\bibinfo {howpublished} {U.S. Patent 4157495} (\bibinfo {year}
  {1979})\BibitemShut {NoStop}%
\bibitem [{\citenamefont {Lam}\ \emph {et~al.}(1983)\citenamefont {Lam},
  \citenamefont {Phillips}, \citenamefont {Kanegsberg},\ and\ \citenamefont
  {Kamin}}]{lam1983}%
  \BibitemOpen
  \bibfield  {author} {\bibinfo {author} {\bibfnamefont {L.~K.}\ \bibnamefont
  {Lam}}, \bibinfo {author} {\bibfnamefont {E.}~\bibnamefont {Phillips}},
  \bibinfo {author} {\bibfnamefont {E.}~\bibnamefont {Kanegsberg}}, \ and\
  \bibinfo {author} {\bibfnamefont {G.~W.}\ \bibnamefont {Kamin}},\ }in\ \href
  {\doibase 10.1117/12.935828} {\emph {\bibinfo {booktitle} {Fiber Optic and
  Laser Sensors I}}},\ Vol.\ \bibinfo {volume} {0412},\ \bibinfo {editor}
  {edited by\ \bibinfo {editor} {\bibfnamefont {E.~L.}\ \bibnamefont {Moore}}\
  and\ \bibinfo {editor} {\bibfnamefont {O.~G.}\ \bibnamefont {Ramer}}},\
  \bibinfo {organization} {International Society for Optics and Photonics}\
  (\bibinfo  {publisher} {SPIE},\ \bibinfo {year} {1983})\ pp.\ \bibinfo
  {pages} {272 -- 276}\BibitemShut {NoStop}%
\bibitem [{\citenamefont {{Larsen}}\ and\ \citenamefont
  {{Bulatowicz}}(2014)}]{larsen2014}%
  \BibitemOpen
  \bibfield  {author} {\bibinfo {author} {\bibfnamefont {M.}~\bibnamefont
  {{Larsen}}}\ and\ \bibinfo {author} {\bibfnamefont {M.}~\bibnamefont
  {{Bulatowicz}}},\ }in\ \href {\doibase 10.1109/ISISS.2014.6782506} {\emph
  {\bibinfo {booktitle} {2014 International Symposium on Inertial Sensors and
  Systems (ISISS)}}}\ (\bibinfo {year} {2014})\ pp.\ \bibinfo {pages}
  {1--5}\BibitemShut {NoStop}%
\bibitem [{\citenamefont {Walker}\ and\ \citenamefont
  {Larsen}(2016)}]{walker16}%
  \BibitemOpen
  \bibfield  {author} {\bibinfo {author} {\bibfnamefont {T.}~\bibnamefont
  {Walker}}\ and\ \bibinfo {author} {\bibfnamefont {M.}~\bibnamefont
  {Larsen}},\ }\href {\doibase https://doi.org/10.1016/bs.aamop.2016.04.002}
  {\bibfield  {journal} {\bibinfo  {journal} {Advances in Atomic Molecular and
  Optical Physics}\ }\textbf {\bibinfo {volume} {65}},\ \bibinfo {pages} {373 }
  (\bibinfo {year} {2016})}\BibitemShut {NoStop}%
\bibitem [{\citenamefont {Korver}\ \emph {et~al.}(2015)\citenamefont {Korver},
  \citenamefont {Thrasher}, \citenamefont {Bulatowicz},\ and\ \citenamefont
  {Walker}}]{kover2015}%
  \BibitemOpen
  \bibfield  {author} {\bibinfo {author} {\bibfnamefont {A.}~\bibnamefont
  {Korver}}, \bibinfo {author} {\bibfnamefont {D.}~\bibnamefont {Thrasher}},
  \bibinfo {author} {\bibfnamefont {M.}~\bibnamefont {Bulatowicz}}, \ and\
  \bibinfo {author} {\bibfnamefont {T.~G.}\ \bibnamefont {Walker}},\ }\href
  {\doibase 10.1103/PhysRevLett.115.253001} {\bibfield  {journal} {\bibinfo
  {journal} {Phys. Rev. Lett.}\ }\textbf {\bibinfo {volume} {115}},\ \bibinfo
  {pages} {253001} (\bibinfo {year} {2015})}\BibitemShut {NoStop}%
\bibitem [{\citenamefont {Thrasher}\ \emph {et~al.}(2019)\citenamefont
  {Thrasher}, \citenamefont {Sorensen}, \citenamefont {Weber}, \citenamefont
  {Bulatowicz}, \citenamefont {Korver}, \citenamefont {Larsen},\ and\
  \citenamefont {Walker}}]{thrasher2019}%
  \BibitemOpen
  \bibfield  {author} {\bibinfo {author} {\bibfnamefont {D.~A.}\ \bibnamefont
  {Thrasher}}, \bibinfo {author} {\bibfnamefont {S.~S.}\ \bibnamefont
  {Sorensen}}, \bibinfo {author} {\bibfnamefont {J.}~\bibnamefont {Weber}},
  \bibinfo {author} {\bibfnamefont {M.}~\bibnamefont {Bulatowicz}}, \bibinfo
  {author} {\bibfnamefont {A.}~\bibnamefont {Korver}}, \bibinfo {author}
  {\bibfnamefont {M.}~\bibnamefont {Larsen}}, \ and\ \bibinfo {author}
  {\bibfnamefont {T.~G.}\ \bibnamefont {Walker}},\ }\href {\doibase
  10.1103/PhysRevA.100.061403} {\bibfield  {journal} {\bibinfo  {journal}
  {Phys. Rev. A}\ }\textbf {\bibinfo {volume} {100}},\ \bibinfo {pages}
  {061403(R)} (\bibinfo {year} {2019})}\BibitemShut {NoStop}%
\bibitem [{\citenamefont {{Slocum}}\ and\ \citenamefont
  {{Marton}}(1973)}]{slocum1973}%
  \BibitemOpen
  \bibfield  {author} {\bibinfo {author} {\bibfnamefont {R.}~\bibnamefont
  {{Slocum}}}\ and\ \bibinfo {author} {\bibfnamefont {B.}~\bibnamefont
  {{Marton}}},\ }\href {\doibase 10.1109/TMAG.1973.1067647} {\bibfield
  {journal} {\bibinfo  {journal} {IEEE Transactions on Magnetics}\ }\textbf
  {\bibinfo {volume} {9}},\ \bibinfo {pages} {221} (\bibinfo {year}
  {1973})}\BibitemShut {NoStop}%
\bibitem [{\citenamefont {Volk}\ \emph {et~al.}(1980)\citenamefont {Volk},
  \citenamefont {Kwon},\ and\ \citenamefont {Mark}}]{volk1980}%
  \BibitemOpen
  \bibfield  {author} {\bibinfo {author} {\bibfnamefont {C.~H.}\ \bibnamefont
  {Volk}}, \bibinfo {author} {\bibfnamefont {T.~M.}\ \bibnamefont {Kwon}}, \
  and\ \bibinfo {author} {\bibfnamefont {J.~G.}\ \bibnamefont {Mark}},\ }\href
  {\doibase 10.1103/PhysRevA.21.1549} {\bibfield  {journal} {\bibinfo
  {journal} {Phys. Rev. A}\ }\textbf {\bibinfo {volume} {21}},\ \bibinfo
  {pages} {1549} (\bibinfo {year} {1980})}\BibitemShut {NoStop}%
\bibitem [{\citenamefont {Li}\ \emph {et~al.}(2006)\citenamefont {Li},
  \citenamefont {Wakai},\ and\ \citenamefont {Walker}}]{Li2006}%
  \BibitemOpen
  \bibfield  {author} {\bibinfo {author} {\bibfnamefont {Z.}~\bibnamefont
  {Li}}, \bibinfo {author} {\bibfnamefont {R.~T.}\ \bibnamefont {Wakai}}, \
  and\ \bibinfo {author} {\bibfnamefont {T.~G.}\ \bibnamefont {Walker}},\
  }\href {\doibase 10.1063/1.2357553} {\bibfield  {journal} {\bibinfo
  {journal} {Appl. Phys. Lett.}\ }\textbf {\bibinfo {volume} {89}},\ \bibinfo
  {pages} {134105} (\bibinfo {year} {2006})}\BibitemShut {NoStop}%
\bibitem [{\citenamefont {Herriott}\ and\ \citenamefont
  {Schulte}(1965)}]{Herriott1965}%
  \BibitemOpen
  \bibfield  {author} {\bibinfo {author} {\bibfnamefont {D.~R.}\ \bibnamefont
  {Herriott}}\ and\ \bibinfo {author} {\bibfnamefont {H.~J.}\ \bibnamefont
  {Schulte}},\ }\href {\doibase 10.1364/AO.4.000883} {\bibfield  {journal}
  {\bibinfo  {journal} {Appl. Opt.}\ }\textbf {\bibinfo {volume} {4}},\
  \bibinfo {pages} {883} (\bibinfo {year} {1965})}\BibitemShut {NoStop}%
\bibitem [{\citenamefont {Silver}(2005)}]{silver2005}%
  \BibitemOpen
  \bibfield  {author} {\bibinfo {author} {\bibfnamefont {J.~A.}\ \bibnamefont
  {Silver}},\ }\href {\doibase 10.1364/AO.44.006545} {\bibfield  {journal}
  {\bibinfo  {journal} {Appl. Opt.}\ }\textbf {\bibinfo {volume} {44}},\
  \bibinfo {pages} {6545} (\bibinfo {year} {2005})}\BibitemShut {NoStop}%
\bibitem [{\citenamefont {Shi}\ \emph {et~al.}(2013)\citenamefont {Shi},
  \citenamefont {Ikäläinen}, \citenamefont {Vaara},\ and\ \citenamefont
  {Romalis}}]{shi2013}%
  \BibitemOpen
  \bibfield  {author} {\bibinfo {author} {\bibfnamefont {J.}~\bibnamefont
  {Shi}}, \bibinfo {author} {\bibfnamefont {S.}~\bibnamefont {Ikäläinen}},
  \bibinfo {author} {\bibfnamefont {J.}~\bibnamefont {Vaara}}, \ and\ \bibinfo
  {author} {\bibfnamefont {M.~V.}\ \bibnamefont {Romalis}},\ }\href {\doibase
  10.1021/jz3018539} {\bibfield  {journal} {\bibinfo  {journal} {J. Phys. Chem.
  Lett.}\ }\textbf {\bibinfo {volume} {4}},\ \bibinfo {pages} {437} (\bibinfo
  {year} {2013})}\BibitemShut {NoStop}%
\bibitem [{\citenamefont {Sheng}\ \emph {et~al.}(2013)\citenamefont {Sheng},
  \citenamefont {Li}, \citenamefont {Dural},\ and\ \citenamefont
  {Romalis}}]{sheng2013}%
  \BibitemOpen
  \bibfield  {author} {\bibinfo {author} {\bibfnamefont {D.}~\bibnamefont
  {Sheng}}, \bibinfo {author} {\bibfnamefont {S.}~\bibnamefont {Li}}, \bibinfo
  {author} {\bibfnamefont {N.}~\bibnamefont {Dural}}, \ and\ \bibinfo {author}
  {\bibfnamefont {M.~V.}\ \bibnamefont {Romalis}},\ }\href {\doibase
  10.1103/PhysRevLett.110.160802} {\bibfield  {journal} {\bibinfo  {journal}
  {Phys. Rev. Lett.}\ }\textbf {\bibinfo {volume} {110}},\ \bibinfo {pages}
  {160802} (\bibinfo {year} {2013})}\BibitemShut {NoStop}%
\bibitem [{\citenamefont {Cooper}\ \emph {et~al.}(2016)\citenamefont {Cooper},
  \citenamefont {Prescott}, \citenamefont {Matz}, \citenamefont {Sauer},
  \citenamefont {Dural}, \citenamefont {Romalis}, \citenamefont {Foley},
  \citenamefont {Kornack}, \citenamefont {Monti},\ and\ \citenamefont
  {Okamitsu}}]{cooper2016}%
  \BibitemOpen
  \bibfield  {author} {\bibinfo {author} {\bibfnamefont {R.~J.}\ \bibnamefont
  {Cooper}}, \bibinfo {author} {\bibfnamefont {D.~W.}\ \bibnamefont
  {Prescott}}, \bibinfo {author} {\bibfnamefont {P.}~\bibnamefont {Matz}},
  \bibinfo {author} {\bibfnamefont {K.~L.}\ \bibnamefont {Sauer}}, \bibinfo
  {author} {\bibfnamefont {N.}~\bibnamefont {Dural}}, \bibinfo {author}
  {\bibfnamefont {M.~V.}\ \bibnamefont {Romalis}}, \bibinfo {author}
  {\bibfnamefont {E.~L.}\ \bibnamefont {Foley}}, \bibinfo {author}
  {\bibfnamefont {T.~W.}\ \bibnamefont {Kornack}}, \bibinfo {author}
  {\bibfnamefont {M.}~\bibnamefont {Monti}}, \ and\ \bibinfo {author}
  {\bibfnamefont {J.}~\bibnamefont {Okamitsu}},\ }\href {\doibase
  10.1103/PhysRevApplied.6.064014} {\bibfield  {journal} {\bibinfo  {journal}
  {Phys. Rev. Applied}\ }\textbf {\bibinfo {volume} {6}},\ \bibinfo {pages}
  {064014} (\bibinfo {year} {2016})}\BibitemShut {NoStop}%
\bibitem [{\citenamefont {Cai}\ \emph {et~al.}(2020)\citenamefont {Cai},
  \citenamefont {Hao}, \citenamefont {Qiu}, \citenamefont {Yu}, \citenamefont
  {Xiao},\ and\ \citenamefont {Sheng}}]{cai2020}%
  \BibitemOpen
  \bibfield  {author} {\bibinfo {author} {\bibfnamefont {B.}~\bibnamefont
  {Cai}}, \bibinfo {author} {\bibfnamefont {C.-P.}\ \bibnamefont {Hao}},
  \bibinfo {author} {\bibfnamefont {Z.-R.}\ \bibnamefont {Qiu}}, \bibinfo
  {author} {\bibfnamefont {Q.-Q.}\ \bibnamefont {Yu}}, \bibinfo {author}
  {\bibfnamefont {W.}~\bibnamefont {Xiao}}, \ and\ \bibinfo {author}
  {\bibfnamefont {D.}~\bibnamefont {Sheng}},\ }\href {\doibase
  10.1103/PhysRevA.101.053436} {\bibfield  {journal} {\bibinfo  {journal}
  {Phys. Rev. A}\ }\textbf {\bibinfo {volume} {101}},\ \bibinfo {pages}
  {053436} (\bibinfo {year} {2020})}\BibitemShut {NoStop}%
\bibitem [{\citenamefont {Limes}\ \emph {et~al.}(2020)\citenamefont {Limes},
  \citenamefont {Foley}, \citenamefont {Kornack}, \citenamefont {Caliga},
  \citenamefont {McBride}, \citenamefont {Braun}, \citenamefont {Lee},
  \citenamefont {Lucivero},\ and\ \citenamefont {Romalis}}]{limes2020}%
  \BibitemOpen
  \bibfield  {author} {\bibinfo {author} {\bibfnamefont {M.}~\bibnamefont
  {Limes}}, \bibinfo {author} {\bibfnamefont {E.}~\bibnamefont {Foley}},
  \bibinfo {author} {\bibfnamefont {T.}~\bibnamefont {Kornack}}, \bibinfo
  {author} {\bibfnamefont {S.}~\bibnamefont {Caliga}}, \bibinfo {author}
  {\bibfnamefont {S.}~\bibnamefont {McBride}}, \bibinfo {author} {\bibfnamefont
  {A.}~\bibnamefont {Braun}}, \bibinfo {author} {\bibfnamefont
  {W.}~\bibnamefont {Lee}}, \bibinfo {author} {\bibfnamefont {V.}~\bibnamefont
  {Lucivero}}, \ and\ \bibinfo {author} {\bibfnamefont {M.}~\bibnamefont
  {Romalis}},\ }\href {\doibase 10.1103/PhysRevApplied.14.011002} {\bibfield
  {journal} {\bibinfo  {journal} {Phys. Rev. Applied}\ }\textbf {\bibinfo
  {volume} {14}},\ \bibinfo {pages} {011002} (\bibinfo {year}
  {2020})}\BibitemShut {NoStop}%
\bibitem [{\citenamefont {Lucivero}\ \emph {et~al.}(2021)\citenamefont
  {Lucivero}, \citenamefont {Lee}, \citenamefont {Dural},\ and\ \citenamefont
  {Romalis}}]{lucivero2021}%
  \BibitemOpen
  \bibfield  {author} {\bibinfo {author} {\bibfnamefont {V.}~\bibnamefont
  {Lucivero}}, \bibinfo {author} {\bibfnamefont {W.}~\bibnamefont {Lee}},
  \bibinfo {author} {\bibfnamefont {N.}~\bibnamefont {Dural}}, \ and\ \bibinfo
  {author} {\bibfnamefont {M.}~\bibnamefont {Romalis}},\ }\href {\doibase
  10.1103/PhysRevApplied.15.014004} {\bibfield  {journal} {\bibinfo  {journal}
  {Phys. Rev. Applied}\ }\textbf {\bibinfo {volume} {15}},\ \bibinfo {pages}
  {014004} (\bibinfo {year} {2021})}\BibitemShut {NoStop}%
\bibitem [{\citenamefont {Kwon}\ \emph {et~al.}(1981)\citenamefont {Kwon},
  \citenamefont {Mark},\ and\ \citenamefont {Volk}}]{kwon81}%
  \BibitemOpen
  \bibfield  {author} {\bibinfo {author} {\bibfnamefont {T.~M.}\ \bibnamefont
  {Kwon}}, \bibinfo {author} {\bibfnamefont {J.~G.}\ \bibnamefont {Mark}}, \
  and\ \bibinfo {author} {\bibfnamefont {C.~H.}\ \bibnamefont {Volk}},\ }\href
  {\doibase 10.1103/PhysRevA.24.1894} {\bibfield  {journal} {\bibinfo
  {journal} {Phys. Rev. A}\ }\textbf {\bibinfo {volume} {24}},\ \bibinfo
  {pages} {1894} (\bibinfo {year} {1981})}\BibitemShut {NoStop}%
\bibitem [{\citenamefont {Li}\ \emph {et~al.}(2011)\citenamefont {Li},
  \citenamefont {Vachaspati}, \citenamefont {Sheng}, \citenamefont {Dural},\
  and\ \citenamefont {Romalis}}]{li2011}%
  \BibitemOpen
  \bibfield  {author} {\bibinfo {author} {\bibfnamefont {S.}~\bibnamefont
  {Li}}, \bibinfo {author} {\bibfnamefont {P.}~\bibnamefont {Vachaspati}},
  \bibinfo {author} {\bibfnamefont {D.}~\bibnamefont {Sheng}}, \bibinfo
  {author} {\bibfnamefont {N.}~\bibnamefont {Dural}}, \ and\ \bibinfo {author}
  {\bibfnamefont {M.~V.}\ \bibnamefont {Romalis}},\ }\href {\doibase
  10.1103/PhysRevA.84.061403} {\bibfield  {journal} {\bibinfo  {journal} {Phys.
  Rev. A}\ }\textbf {\bibinfo {volume} {84}},\ \bibinfo {pages} {061403}
  (\bibinfo {year} {2011})}\BibitemShut {NoStop}%
\bibitem [{\citenamefont {Libbrecht}\ and\ \citenamefont
  {Hall}(1993)}]{libbrecht1993}%
  \BibitemOpen
  \bibfield  {author} {\bibinfo {author} {\bibfnamefont {K.~G.}\ \bibnamefont
  {Libbrecht}}\ and\ \bibinfo {author} {\bibfnamefont {J.~L.}\ \bibnamefont
  {Hall}},\ }\href {\doibase 10.1063/1.1143949} {\bibfield  {journal} {\bibinfo
   {journal} {Rev. Sci. Instrum.}\ }\textbf {\bibinfo {volume} {64}},\ \bibinfo
  {pages} {2133} (\bibinfo {year} {1993})}\BibitemShut {NoStop}%
\bibitem [{\citenamefont {Zhao}\ \emph {et~al.}(1998)\citenamefont {Zhao},
  \citenamefont {Simsarian}, \citenamefont {Orozco},\ and\ \citenamefont
  {Sprouse}}]{zhao1998}%
  \BibitemOpen
  \bibfield  {author} {\bibinfo {author} {\bibfnamefont {W.~Z.}\ \bibnamefont
  {Zhao}}, \bibinfo {author} {\bibfnamefont {J.~E.}\ \bibnamefont {Simsarian}},
  \bibinfo {author} {\bibfnamefont {L.~A.}\ \bibnamefont {Orozco}}, \ and\
  \bibinfo {author} {\bibfnamefont {G.~D.}\ \bibnamefont {Sprouse}},\ }\href
  {\doibase 10.1063/1.1149171} {\bibfield  {journal} {\bibinfo  {journal} {Rev.
  Sci. Instrum.}\ }\textbf {\bibinfo {volume} {69}},\ \bibinfo {pages} {3737}
  (\bibinfo {year} {1998})}\BibitemShut {NoStop}%
\bibitem [{\citenamefont {Smullin}\ \emph {et~al.}(2009)\citenamefont
  {Smullin}, \citenamefont {Savukov}, \citenamefont {Vasilakis}, \citenamefont
  {Ghosh},\ and\ \citenamefont {Romalis}}]{smullin2009}%
  \BibitemOpen
  \bibfield  {author} {\bibinfo {author} {\bibfnamefont {S.~J.}\ \bibnamefont
  {Smullin}}, \bibinfo {author} {\bibfnamefont {I.~M.}\ \bibnamefont
  {Savukov}}, \bibinfo {author} {\bibfnamefont {G.}~\bibnamefont {Vasilakis}},
  \bibinfo {author} {\bibfnamefont {R.~K.}\ \bibnamefont {Ghosh}}, \ and\
  \bibinfo {author} {\bibfnamefont {M.~V.}\ \bibnamefont {Romalis}},\ }\href
  {\doibase 10.1103/PhysRevA.80.033420} {\bibfield  {journal} {\bibinfo
  {journal} {Phys. Rev. A}\ }\textbf {\bibinfo {volume} {80}},\ \bibinfo
  {pages} {033420} (\bibinfo {year} {2009})}\BibitemShut {NoStop}%
\bibitem [{\citenamefont {Rubiola}(2009)}]{rubiola2009}%
  \BibitemOpen
  \bibfield  {author} {\bibinfo {author} {\bibfnamefont {E.}~\bibnamefont
  {Rubiola}},\ }\href@noop {} {\emph {\bibinfo {title} {Phase noise and
  Frequency Stability in Oscillators}}}\ (\bibinfo  {publisher} {Cambridge
  University Press},\ \bibinfo {address} {Cambridge, UK},\ \bibinfo {year}
  {2009})\BibitemShut {NoStop}%
\bibitem [{\citenamefont {Zhang}\ \emph {et~al.}(2020)\citenamefont {Zhang},
  \citenamefont {Zhao},\ and\ \citenamefont {Wang}}]{zhang2020}%
  \BibitemOpen
  \bibfield  {author} {\bibinfo {author} {\bibfnamefont {K.}~\bibnamefont
  {Zhang}}, \bibinfo {author} {\bibfnamefont {N.}~\bibnamefont {Zhao}}, \ and\
  \bibinfo {author} {\bibfnamefont {Y.-H.}\ \bibnamefont {Wang}},\ }\href
  {\doibase 10.1038/s41598-020-59088-y} {\bibfield  {journal} {\bibinfo
  {journal} {Sci. Rep.}\ }\textbf {\bibinfo {volume} {10}},\ \bibinfo {pages}
  {2258} (\bibinfo {year} {2020})}\BibitemShut {NoStop}%
\bibitem [{\citenamefont {Pospelov}\ \emph {et~al.}(2013)\citenamefont
  {Pospelov}, \citenamefont {Pustelny}, \citenamefont {Ledbetter},
  \citenamefont {Kimball}, \citenamefont {Gawlik},\ and\ \citenamefont
  {Budker}}]{pospelov2013}%
  \BibitemOpen
  \bibfield  {author} {\bibinfo {author} {\bibfnamefont {M.}~\bibnamefont
  {Pospelov}}, \bibinfo {author} {\bibfnamefont {S.}~\bibnamefont {Pustelny}},
  \bibinfo {author} {\bibfnamefont {M.~P.}\ \bibnamefont {Ledbetter}}, \bibinfo
  {author} {\bibfnamefont {D.~F.~J.}\ \bibnamefont {Kimball}}, \bibinfo
  {author} {\bibfnamefont {W.}~\bibnamefont {Gawlik}}, \ and\ \bibinfo {author}
  {\bibfnamefont {D.}~\bibnamefont {Budker}},\ }\href {\doibase
  10.1103/PhysRevLett.110.021803} {\bibfield  {journal} {\bibinfo  {journal}
  {Phys. Rev. Lett.}\ }\textbf {\bibinfo {volume} {110}},\ \bibinfo {pages}
  {021803} (\bibinfo {year} {2013})}\BibitemShut {NoStop}%
\bibitem [{\citenamefont {Pustelny}\ \emph {et~al.}(2013)\citenamefont
  {Pustelny}, \citenamefont {Kimball}, \citenamefont {Pankow}, \citenamefont
  {Ledbetter}, \citenamefont {Wlodarczyk}, \citenamefont {Wcislo},
  \citenamefont {Pospelov}, \citenamefont {Smith}, \citenamefont {Read},
  \citenamefont {Gawlik} \emph {et~al.}}]{pustelny2013}%
  \BibitemOpen
  \bibfield  {author} {\bibinfo {author} {\bibfnamefont {S.}~\bibnamefont
  {Pustelny}}, \bibinfo {author} {\bibfnamefont {D.~F.}\ \bibnamefont
  {Kimball}}, \bibinfo {author} {\bibfnamefont {C.}~\bibnamefont {Pankow}},
  \bibinfo {author} {\bibfnamefont {M.~P.}\ \bibnamefont {Ledbetter}}, \bibinfo
  {author} {\bibfnamefont {P.}~\bibnamefont {Wlodarczyk}}, \bibinfo {author}
  {\bibfnamefont {P.}~\bibnamefont {Wcislo}}, \bibinfo {author} {\bibfnamefont
  {M.}~\bibnamefont {Pospelov}}, \bibinfo {author} {\bibfnamefont {J.~R.}\
  \bibnamefont {Smith}}, \bibinfo {author} {\bibfnamefont {J.}~\bibnamefont
  {Read}}, \bibinfo {author} {\bibfnamefont {W.}~\bibnamefont {Gawlik}},  \emph
  {et~al.},\ }\href@noop {} {\bibfield  {journal} {\bibinfo  {journal} {Ann.
  Phys.}\ }\textbf {\bibinfo {volume} {525}},\ \bibinfo {pages} {659} (\bibinfo
  {year} {2013})}\BibitemShut {NoStop}%
\bibitem [{\citenamefont {Masia-Roig}\ \emph {et~al.}(2020)\citenamefont
  {Masia-Roig}, \citenamefont {Smiga}, \citenamefont {Budker}, \citenamefont
  {Dumont}, \citenamefont {Grujic}, \citenamefont {Kim}, \citenamefont
  {{Jackson Kimball}}, \citenamefont {Lebedev}, \citenamefont {Monroy},
  \citenamefont {Pustelny}, \citenamefont {Scholtes}, \citenamefont {Segura},
  \citenamefont {Semertzidis}, \citenamefont {Shin}, \citenamefont {Stalnaker},
  \citenamefont {Sulai}, \citenamefont {Weis},\ and\ \citenamefont
  {Wickenbrock}}]{hector2020}%
  \BibitemOpen
  \bibfield  {author} {\bibinfo {author} {\bibfnamefont {H.}~\bibnamefont
  {Masia-Roig}}, \bibinfo {author} {\bibfnamefont {J.~A.}\ \bibnamefont
  {Smiga}}, \bibinfo {author} {\bibfnamefont {D.}~\bibnamefont {Budker}},
  \bibinfo {author} {\bibfnamefont {V.}~\bibnamefont {Dumont}}, \bibinfo
  {author} {\bibfnamefont {Z.}~\bibnamefont {Grujic}}, \bibinfo {author}
  {\bibfnamefont {D.}~\bibnamefont {Kim}}, \bibinfo {author} {\bibfnamefont
  {D.~F.}\ \bibnamefont {{Jackson Kimball}}}, \bibinfo {author} {\bibfnamefont
  {V.}~\bibnamefont {Lebedev}}, \bibinfo {author} {\bibfnamefont
  {M.}~\bibnamefont {Monroy}}, \bibinfo {author} {\bibfnamefont
  {S.}~\bibnamefont {Pustelny}}, \bibinfo {author} {\bibfnamefont
  {T.}~\bibnamefont {Scholtes}}, \bibinfo {author} {\bibfnamefont {P.~C.}\
  \bibnamefont {Segura}}, \bibinfo {author} {\bibfnamefont {Y.~K.}\
  \bibnamefont {Semertzidis}}, \bibinfo {author} {\bibfnamefont {Y.~C.}\
  \bibnamefont {Shin}}, \bibinfo {author} {\bibfnamefont {J.~E.}\ \bibnamefont
  {Stalnaker}}, \bibinfo {author} {\bibfnamefont {I.}~\bibnamefont {Sulai}},
  \bibinfo {author} {\bibfnamefont {A.}~\bibnamefont {Weis}}, \ and\ \bibinfo
  {author} {\bibfnamefont {A.}~\bibnamefont {Wickenbrock}},\ }\href {\doibase
  https://doi.org/10.1016/j.dark.2020.100494} {\bibfield  {journal} {\bibinfo
  {journal} {Phys. Dark Universe}\ }\textbf {\bibinfo {volume} {28}},\ \bibinfo
  {pages} {100494} (\bibinfo {year} {2020})}\BibitemShut {NoStop}%
\bibitem [{\citenamefont {Afach}\ \emph {et~al.}(2021)\citenamefont {Afach},
  \citenamefont {Buchler}, \citenamefont {Budker}, \citenamefont {Dailey},
  \citenamefont {Derevianko}, \citenamefont {Dumont}, \citenamefont {Figueroa},
  \citenamefont {Gerhardt}, \citenamefont {Grujić}, \citenamefont {Guo},
  \citenamefont {Hao}, \citenamefont {Hamilton}, \citenamefont {Hedges},
  \citenamefont {Kimball}, \citenamefont {Kim}, \citenamefont {Khamis},
  \citenamefont {Kornack}, \citenamefont {Lebedev}, \citenamefont {Lu},
  \citenamefont {Masia-Roig}, \citenamefont {Monroy}, \citenamefont {Padniuk},
  \citenamefont {Palm}, \citenamefont {Park}, \citenamefont {Paul},
  \citenamefont {Penaflor}, \citenamefont {Peng}, \citenamefont {Pospelov},
  \citenamefont {Preston}, \citenamefont {Pustelny}, \citenamefont {Scholtes},
  \citenamefont {Segura}, \citenamefont {Semertzidis}, \citenamefont {Sheng},
  \citenamefont {Shin}, \citenamefont {Smiga}, \citenamefont {Stalnaker},
  \citenamefont {Sulai}, \citenamefont {Tandon}, \citenamefont {Wang},
  \citenamefont {Weis}, \citenamefont {Wickenbrock}, \citenamefont {Wilson},
  \citenamefont {Wu}, \citenamefont {Wurm}, \citenamefont {Xiao}, \citenamefont
  {Yang}, \citenamefont {Yu},\ and\ \citenamefont {Zhang}}]{afach2021}%
  \BibitemOpen
  \bibfield  {author} {\bibinfo {author} {\bibfnamefont {S.}~\bibnamefont
  {Afach}}, \bibinfo {author} {\bibfnamefont {B.~C.}\ \bibnamefont {Buchler}},
  \bibinfo {author} {\bibfnamefont {D.}~\bibnamefont {Budker}}, \bibinfo
  {author} {\bibfnamefont {C.}~\bibnamefont {Dailey}}, \bibinfo {author}
  {\bibfnamefont {A.}~\bibnamefont {Derevianko}}, \bibinfo {author}
  {\bibfnamefont {V.}~\bibnamefont {Dumont}}, \bibinfo {author} {\bibfnamefont
  {N.~L.}\ \bibnamefont {Figueroa}}, \bibinfo {author} {\bibfnamefont
  {I.}~\bibnamefont {Gerhardt}}, \bibinfo {author} {\bibfnamefont {Z.~D.}\
  \bibnamefont {Grujić}}, \bibinfo {author} {\bibfnamefont {H.}~\bibnamefont
  {Guo}}, \bibinfo {author} {\bibfnamefont {C.}~\bibnamefont {Hao}}, \bibinfo
  {author} {\bibfnamefont {P.~S.}\ \bibnamefont {Hamilton}}, \bibinfo {author}
  {\bibfnamefont {M.}~\bibnamefont {Hedges}}, \bibinfo {author} {\bibfnamefont
  {D.~F.~J.}\ \bibnamefont {Kimball}}, \bibinfo {author} {\bibfnamefont
  {D.}~\bibnamefont {Kim}}, \bibinfo {author} {\bibfnamefont {S.}~\bibnamefont
  {Khamis}}, \bibinfo {author} {\bibfnamefont {T.}~\bibnamefont {Kornack}},
  \bibinfo {author} {\bibfnamefont {V.}~\bibnamefont {Lebedev}}, \bibinfo
  {author} {\bibfnamefont {Z.-T.}\ \bibnamefont {Lu}}, \bibinfo {author}
  {\bibfnamefont {H.}~\bibnamefont {Masia-Roig}}, \bibinfo {author}
  {\bibfnamefont {M.}~\bibnamefont {Monroy}}, \bibinfo {author} {\bibfnamefont
  {M.}~\bibnamefont {Padniuk}}, \bibinfo {author} {\bibfnamefont {C.~A.}\
  \bibnamefont {Palm}}, \bibinfo {author} {\bibfnamefont {S.~Y.}\ \bibnamefont
  {Park}}, \bibinfo {author} {\bibfnamefont {K.~V.}\ \bibnamefont {Paul}},
  \bibinfo {author} {\bibfnamefont {A.}~\bibnamefont {Penaflor}}, \bibinfo
  {author} {\bibfnamefont {X.}~\bibnamefont {Peng}}, \bibinfo {author}
  {\bibfnamefont {M.}~\bibnamefont {Pospelov}}, \bibinfo {author}
  {\bibfnamefont {R.}~\bibnamefont {Preston}}, \bibinfo {author} {\bibfnamefont
  {S.}~\bibnamefont {Pustelny}}, \bibinfo {author} {\bibfnamefont
  {T.}~\bibnamefont {Scholtes}}, \bibinfo {author} {\bibfnamefont {P.~C.}\
  \bibnamefont {Segura}}, \bibinfo {author} {\bibfnamefont {Y.~K.}\
  \bibnamefont {Semertzidis}}, \bibinfo {author} {\bibfnamefont
  {D.}~\bibnamefont {Sheng}}, \bibinfo {author} {\bibfnamefont {Y.~C.}\
  \bibnamefont {Shin}}, \bibinfo {author} {\bibfnamefont {J.~A.}\ \bibnamefont
  {Smiga}}, \bibinfo {author} {\bibfnamefont {J.~E.}\ \bibnamefont
  {Stalnaker}}, \bibinfo {author} {\bibfnamefont {I.}~\bibnamefont {Sulai}},
  \bibinfo {author} {\bibfnamefont {D.}~\bibnamefont {Tandon}}, \bibinfo
  {author} {\bibfnamefont {T.}~\bibnamefont {Wang}}, \bibinfo {author}
  {\bibfnamefont {A.}~\bibnamefont {Weis}}, \bibinfo {author} {\bibfnamefont
  {A.}~\bibnamefont {Wickenbrock}}, \bibinfo {author} {\bibfnamefont
  {T.}~\bibnamefont {Wilson}}, \bibinfo {author} {\bibfnamefont
  {T.}~\bibnamefont {Wu}}, \bibinfo {author} {\bibfnamefont {D.}~\bibnamefont
  {Wurm}}, \bibinfo {author} {\bibfnamefont {W.}~\bibnamefont {Xiao}}, \bibinfo
  {author} {\bibfnamefont {Y.}~\bibnamefont {Yang}}, \bibinfo {author}
  {\bibfnamefont {D.}~\bibnamefont {Yu}}, \ and\ \bibinfo {author}
  {\bibfnamefont {J.}~\bibnamefont {Zhang}},\ }\href@noop {} {} (\bibinfo
  {year} {2021}),\ \Eprint {http://arxiv.org/abs/2102.13379} {arXiv:2102.13379
  [astro-ph.CO]} \BibitemShut {NoStop}%
\end{thebibliography}
\end{document}